\documentclass[aps,twocolumn,superscriptaddress,showpacs,amssymb,prb,floatfix,preprintnumbers]{revtex4-1}
\usepackage{graphicx,color}
\usepackage{dcolumn}
\usepackage{amsmath}
\usepackage{amssymb}
\usepackage[dvipdfm]{hyperref}
\hypersetup{colorlinks=true,linkcolor=blue,
  implicit=true,breaklinks=true,pagebackref=true,backref=true,
  bookmarks=true,bookmarksnumbered=true,hyperfootnotes=true,debug=true,
  naturalnames=false,citecolor=blue,pdfview=FitH,pdfstartview=FitH,hyperindex=true}

\newcommand{\slrrtext}  {spin-lattice-relaxation rate}

\newcommand{\bscco}     {Bi$_2$Sr$_2$CaCu$_2$O$_{8+\delta}$}

\newcommand{\etal}      {{\it et al}.}
\newcommand{\slrr}      {$T_1^{-1}$}

\newcommand{\ybcox}    {${\rm Y} {\rm Ba}_{2} {\rm Cu}_{3} {\rm O}_{7-\delta}$}

\newcommand{\lscox}     {${\rm La}_{2-x} {\rm Sr}_{x} {\rm Cu O_4}$}

\newcommand{\oxy}       {$^{17}\!$O}

\begin{document}

\thispagestyle{myheadings}

\title{NMR Studies of pseudogap and electronic inhomogeneity in Bi$_2$Sr$_2$CaCu$_2$O$_{8+\delta}$}

\author{J. Crocker}
\author{A. P. Dioguardi}
\author{N. apRoberts-Warren}
\author{A. C. Shockley}
\affiliation{Department of Physics, University of California, Davis, CA 95616, USA}
\author{H.-J. Grafe}
\affiliation{IFW Dresden, Institute for Solid State Research, P.O. Box 270116, D-01171 Dresden, Germany}
\author{Z. Xu}
\author{J. Wen}
\author{G. Gu}
\affiliation{Condensed Matter Physics and Materials Sciences Department, Brookhaven National Laboratory, Upton, New York 11973, USA}
\author{N. J. Curro}
\email{curro@physics.ucdavis.edu}
\affiliation{Department of Physics, University of California, Davis, CA 95616, USA}


\date{\today}

\begin{abstract}
We report $^{17}$O NMR measurements in single crystals of overdoped Bi$_2$Sr$_2$CaCu$_2$O$_{8+\delta}$ with $T_c=82$ K.  We measure the full anisotropy of the planar oxygen Knight shift, electric field gradient, and spin lattice relaxation rate tensors, and show that the entire temperature dependence is determined by the suppression of the density of states in the pseudogap below $T^*\sim 94$ K.  The linewidth can be explained by a combination of magnetic and quadrupolar  broadening as a result of an inhomogeneous distribution of local hole concentrations that is consistent with scanning tunneling microscopy measurements.  This distribution is temperature independent, does not break $C_4$ symmetry, and exhibits no change below $T^*$ or $T_c$.

\end{abstract}


\maketitle

\section{Introduction}

The normal state of the high temperature superconductors continues to attract attention despite more than twenty years of research. The parent state of these materials is a Mott insulator, but for sufficient hole doping the conductivity becomes metallic and high temperature superconductivity emerges with $T_c \sim 100$ K and d-wave symmetry.\cite{Kivelson1998,MonthouxPinesReview} Other strongly correlated electron systems such as the heavy fermions, the iron-based superconductors
and  the organic superconductors also exhibit superconductivity in close proximity to an antiferromagnetism, \cite{jdtreview,Lefebvre:2000vn,doping122review} but the cuprates are unique in that they also exhibit a pseudogap in the normal state over a broad range of dopings.\cite{TimuskPGreview}   It remains unclear whether the partial suppression of the low energy density of states in the pseudogap is the result of a new thermodynamic phase \cite{StanfordScienceBi2201PG2011,HiddenOrderCupratesPRB2001,VarmaPG1997} or a crossover in magnetic behavior driven by the proximate Mott insulating state.\cite{AndersonPG2006} Of particular interest is the fact that the electronic degrees of freedom in the pseudogap develop short range inhomogeneous mesoscopic structures that break the $C_4$ structural symmetry of the CuO$_2$ plane.\cite{davisnatureinhomogeneity,mcElroySTMScience2005,Lawler2010} Inelastic neutron scattering (INS) studies of these materials reveal magnetic excitations that can be described by one-dimensional stripe structures consistent with the STM results.\cite{tranquadastripesnature,Xu2009}  These inhomogeneous electronic structures may be a more general manifestations of electronic liquid crystalline phases of doped Mott insulators.\cite{Kivelson1998,DavisSTMscience2011}


\begin{figure}
\begin{center}
 \includegraphics[width=\linewidth]{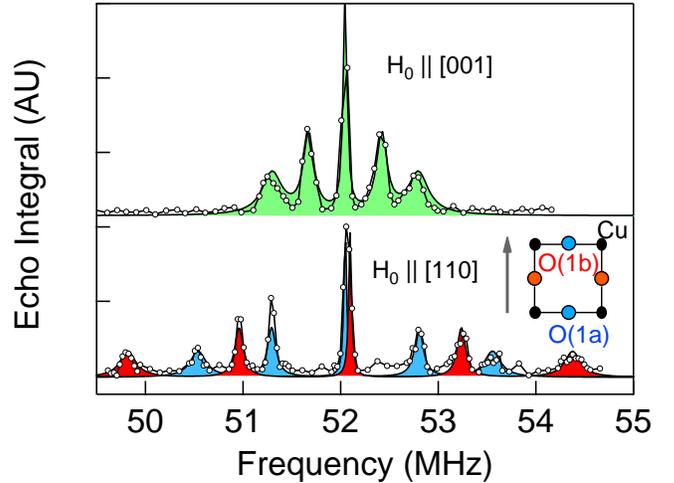}
 \caption{(Color online) Single crystal $^{17}$O NMR spectra of \bscco\ in 9 T for fields $\mathbf{H}_0$ applied along the $c-$axis and in the $ab$-plane at 110 K. Solid lines are fits as described in the text.}
\label{spectra}
\end{center}
\end{figure}

Nuclear magnetic resonance (NMR) of the oxygen nuclei in the cuprates offers unique insight into the interplay of charge and spin degrees of freedom.\cite{GrafeLESCO,Haase2000,JulienGlassyStripes2001PRB,TakigawaONMRinYBCO,MartindaleHammelYBCO,shastrycuprates}
Unlike STM, NMR probes the bulk of the material, and both the quadrupolar moment and nuclear magnetic moment of the \oxy\ couple to the local charge and spin environments, respectively.  \oxy\ ($I=\frac{5}{2}$) NMR can shed light on the spatial correlations between the charge and spin degrees of freedom of these mesoscopic liquid crystalline phases.\cite{GrafeLESCO}    In order to investigate the local environment at the oxygen site, we have measured NMR spectra and the nuclear \slrrtext\ (\slrr) of a single overdoped crystal of \bscco\ ($T_c=82$ K) isotopically enriched with $^{17}$O.  This compound has been studied extensively via surface probes such as STM and angle resolved photoemission (ARPES) because as it cleaves easily revealing well formed surfaces normal to the $c-$direction.\cite{davisnatureinhomogeneity,ZXreview}  On the other hand, \bscco\ is one of the most difficult cuprate families to investigate with NMR because a structural superlattice modulation gives rise to a quadrupolar broadening of the $^{63,65}$Cu and $^{209}$Bi resonances.\cite{hirschfeldBSCCOmodulation,walstedtBSCCO}   With the exception of a few key experiments, there have been no systematic studies of this compound with NMR. Walstedt \etal\ and later Ishida \etal\ reported Cu Knight shift and \slrr\ measurements of pseudogap behavior in underdoped and overdoped \bscco\ crystals;\cite{walstedtBSCCO,IshidaBSCCO} Takigawa and Mitzi reported oxygen and copper NMR evidence for d-wave pairing in the superconducting state; \cite{TakigawaBSCCOprl}  and  recently Chen \etal\ reported evidence for magnetic impurities based on \oxy\ NMR in \bscco.\cite{HalperinBSCCOimpurities}  Each of these studies focused on the spin susceptibility measured at the Cu and/or O sites.  In this paper, in addition to the temperature dependence of the Knight shift, we focus also on the temperature dependence of the electric field gradient (EFG) and \slrr\ tensors at the planar oxygen site in order to investigate the possibility of $C_4$ symmetry breaking and the influence of the inhomogeneous electronic states on the NMR properties.

\section{Spectral Measurements}

The \bscco\ single crystal was grown by a floating zone method.\cite{bsccogrowth}  The superconducting transition temperature of the as-grown single crystal was 91K.  Single crystals were enriched with \oxy\ by annealing for 144 hours at 550 K in an isotopically enriched oxygen atmosphere.  Unlike other high $T_c$ materials, oxygen exchanges quickly in \bscco\ so that the \oxy\ NMR signal is detectable even in a single crystal.  In other cuprates oxygen NMR is usually only possible in aligned samples, in which a powder sample (with greater surface area) is isotopically enriched, then mixed with epoxy and cured in an external field.\cite{GrafeLESCO,MartindaleHammelYBCO,GrafeLBCOxtalNMR}  Aligned powder samples have well resolved spectra for the applied field along the alignment  axis (usually the $c$-direction), but exhibit a random powder pattern for perpendicular fields.\cite{YoshinariYBCO}  Although there are ways to extract parameters of the EFG and \slrr\ tensors from these powder patterns, there are usually unavoidable baseline and orientation distribution corrections. \cite{MartindaleHammelYBCO,curroNMRY1248} On the other hand, isotopically enriched \bscco\ single crystals enable one to  measure directly the full shift, quadrupolar and \slrrtext\ tensors.  This information is crucial because the planar oxygen does not have axial symmetry, and the principal axes of these tensors lie along the Cu-O bond directions and the $c$-axis of the crystal.\cite{TakigawaBSCCOprl,WalstedtPRL1998}.  Measurements of these tensors enable one to extract the full anisotropy of the spin fluctuations and EFG at this site.

\begin{figure}
\begin{center}
 \includegraphics[width=\linewidth]{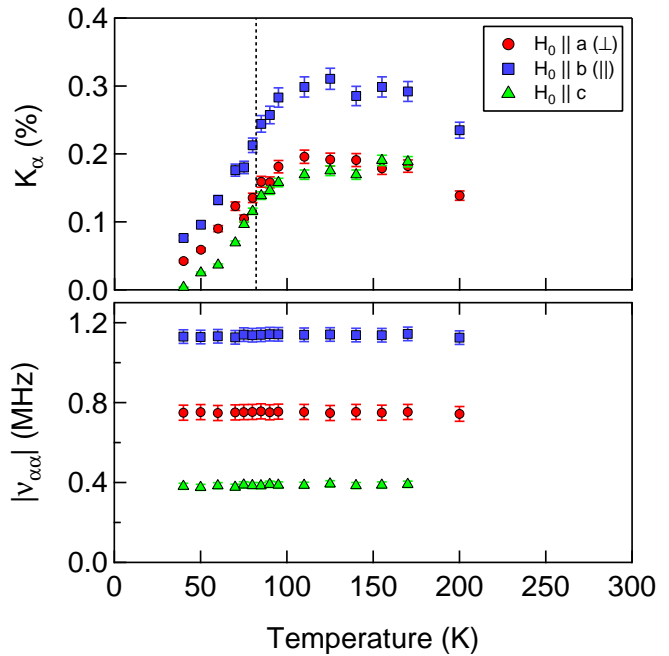}
 \caption{(Color online) The magnetic shift, $K_{\alpha}$ (upper panel) and the electric field gradient $\nu_{\alpha}$ (lower panel) as a function of temperature for $\alpha = a,b,c$.  The colors correspond to the same field orientations as in Fig. \ref{spectra}.  The dotted line in the upper panel indicated $T_c$.}
\label{KandEFGvsT}
\end{center}
\end{figure}

The NMR measurements were carried out using a Quantum Design Physical Property Measurement System (PPMS) in a field of 9 T using a home-built probe, and magnetization measurements were carried out using a Quantum Design SQUID magnetometer. Spectra were acquired by accumulating spin echoes at fixed field, and the \slrrtext\ was measured by inversion recovery.  The Hamiltonian for the planar O(1) nuclei is given by:
\begin{equation}
\label{eqn:hamiltonian}
\mathcal{H} = \gamma\hbar H_0 (1+K_{\alpha})\hat{I}_{\alpha} + \frac{h \nu_{bb}}{6}[(3\hat{I}_b^2 - \hat{I}^2) +
\eta(\hat{I}_c^2-\hat{I}_a^2)],
\end{equation}
where $\gamma= 0.57719$ MHz/T is the gyromagnetic ratio for \oxy, $\hat{I}_{\alpha}$ and $K_{\alpha}$ are the nuclear spin operator and the magnetic shift along the $\alpha$ direction, $H_0$ is the external field, $\nu_{\alpha\alpha}$ are the components of the EFG tensor along the principal axes, and $\eta
=(|V_{cc}|-|V_{aa}|)/|V_{bb}|$ is the asymmetry parameter of the EFG.  We adopt a convention in which the principal axes lie along $a\parallel[110]$ in the CuO$_2$ plane perpendicular to the Cu-O bond axis, $b \parallel [1\bar{1}0]$ is parallel to the Cu-O bond axis (see diagram in Fig. \ref{spectra}), and $c\parallel [001]$ lies perpendicular to the CuO$_2$ plane.   The largest eigenvalue of the EFG tensor  ($\nu_{bb}$) is associated with the bond axis. The EFG components are given by $\nu_{\alpha\alpha} ={3eQV_{\alpha\alpha}}/{20h}$, where $Q=25.6$ mbarn is the quadrupolar moment of the \oxy, and $V_{\alpha\alpha}$ is the second derivative of the potential $V(\mathbf{r})$ along the $\alpha$ direction at the O(1) site.

\begin{figure}
\begin{center}
 \includegraphics[width=\linewidth]{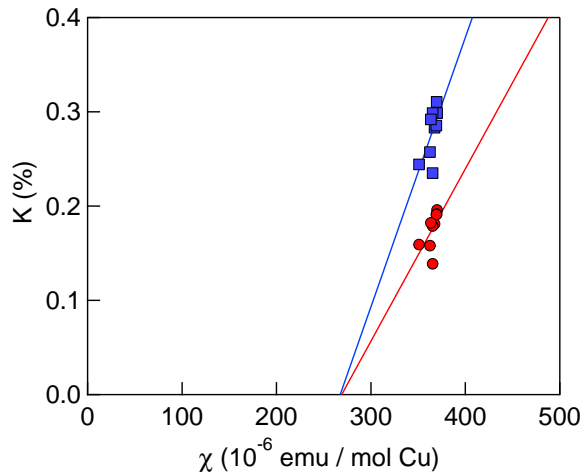}
 \caption{(Color online) $K_{\alpha}$ versus $\chi_{\alpha}$ for $\alpha = a,b$.  The colors match those in Fig. \ref{spectra}.  As described in the text, the best fits to the data give the hyperfine coupling constants $2C_{aa} = 102(40)$ kOe/$\mu_B$ and $2C_{bb}= 160(40)$ kOe/$\mu_B$, and $K_{0a} = -0.5(4)\%$ and $K_{0b} = -0.8(4)\%$.}
\label{Kvschi}
\end{center}
\end{figure}

Fig. \ref{spectra} shows spectra of the oxygen for the field both parallel and perpendicular to the $c$ axis.  Note that when $\mathbf{H}_0$ lies in the plane, the two planar O(1) sites are no longer equivalent, and we define O(1b) as the site with the field parallel to the bond axis, and O(1a) site as the site with the field perpendicular. We do not observe the oxygen site in the SrO layer because \slrr\ for these sites is much longer than the experimental repetition rate. Furthermore, structural disorder from the supermodulation may broaden the oxygen site in the BiO layer as well.\cite{TakigawaBSCCOprl} In order to extract the Knight shift and EFG tensors, we fit the spectra in each direction to a sum of five Gaussians. The second moment of each transition is given by $\sigma = \sqrt{\sigma_M^2 + n\sigma_Q^2}$, where $\sigma_M$ is the rms second moment of magnetic contribution to the linewidth, $\sigma_Q$ is the rms second moment of the quadrupolar contribution to the linewidth, and $n = 0,\pm 1, \pm 2$ indicates the particular nuclear transition.  At 110 K we find $(K_a, K_b, K_c) = (0.19(1), 0.29(1), 0.19(1))$\%, and $(\nu_{aa}, \nu_{bb}, \nu_{cc}) = (-754(3), 1139(4), -375(4))$ kHz (the asymmetry parameter $\eta = 0.33$). These values are close to those reported previously in a crystal with similar $T_c$. \cite{TakigawaBSCCOprl} The EFG arises both from on-site contributions from the unfilled 2p orbitals of the oxygen, and lattice contributions from distant charges.  The dominant contribution arises from the on-site term, and is a direct measure of the doping in the cuprates \cite{HaaseHoleDensity,GrafeLESCO}.
 Susceptibility and resistivity measurements indicate $T_c = 82$ K corresponding to a doping level of $p\approx 0.21$. \cite{MomonoBSCCOgaps}


Figure \ref{KandEFGvsT} shows the temperature dependence of $K_{\alpha}$ and $\nu_{\alpha\alpha}$.    The EFG tensor is temperature independent, suggesting the absence of any static charge order.\cite{GrafeLESCO}   In contrast the magnetic shift is strongly temperature dependent. The magnetic shift is given by $K_{\alpha}(T) = K^{o}_{\alpha} + K_{\alpha}^{dia} + K_{\alpha}^{s}(T)$, where $K^{o}_{\alpha}$, $K^{dia}_{\alpha}$, and $K^{s}_{\alpha}$ are the contributions from the orbital (Van Vleck) susceptibility of the O $2p$ orbitals, $\chi^{o}_{\alpha}$;  the diamagnetic susceptibility of the filled orbitals, $\chi^{dia}_{\alpha}$; and the spin susceptibility of the Cu $S=1/2$ spins, $\chi^{s}_{\alpha}(T)$.  The first two terms are temperature independent and combine to give an overall shift $K_0$.  The spin shift is arises because of the hyperfine coupling to the Cu electron spins:
\begin{equation}
\mathcal{H}_{\rm hyp} = \gamma\hbar g\mu_B\sum_{\mathbf{r}_i} \hat{\mathbf{I}}\cdot\mathbb{C}\cdot\mathbf{S}(\mathbf{r}+\mathbf{r}_i),
\label{eqn:milarice}
\end{equation}
where the sum is over the two nearest neighbor Cu sites, and $\mathbb{C}$ is the transferred hyperfine coupling tensor. \cite{MilaRiceHamiltonian}  The spin shift is given by $K^{s}_{\alpha} = 2C_{\alpha\alpha}\chi^{s}_{\alpha}(T)$, and the coupling parameters  $C_{\alpha\alpha}$ can be determined by the slope of  $K_{\alpha}$ versus $\chi_{ab}$ along the principal axes of the hyperfine tensor in the normal state (see Fig. \ref{Kvschi}). We find $C_{aa} = 51$ kOe/$\mu_B$ and $C_{bb} = 80$ kOe/$\mu_B$, similar to previous measurements in \ybcox.\cite{YoshinariYBCO}  For the $c$-direction, $\chi_c$ is essentially temperature independent, and we do not have enough precision to measure $C_{cc}$; however \slrrtext\ measurements discussed below suggest that $C_{cc}\sim C_{aa}$, in agreement with previous reports in \ybcox.\cite{YoshinariYBCO}  The approximate axial symmetry of the transferred hyperfine coupling arises from the O 2p-orbitals that hybridize with the nearest neighbor Cu 4s orbitals.\cite{MilaRiceHamiltonian}

The pseudogap is manifest in NMR  as a peak in the Knight shift at a temperature $T^m$ followed by a breakdown of scaling below a temperature $T^*$.\cite{BarzykinCuprateScaling} As seen in Fig. \ref{KandEFGvsT}, we find that $K_{\alpha}(T)$  reaches a maximum at a temperature $T^m\approx 125$ K and is suppressed at lower temperatures.\cite{curroNMRY1248,BarzykinCuprateScaling}  This value agrees with other probes of the pseudogap temperature for this doping level in \bscco.\cite{TakigawaBSCCOprl,MomonoBSCCOgaps} $K_{\alpha}$ continues to be suppressed below $T_c$ due to the spin singlet nature of the Cooper pairs.

\begin{figure}
\begin{center}
 \includegraphics[width=\linewidth]{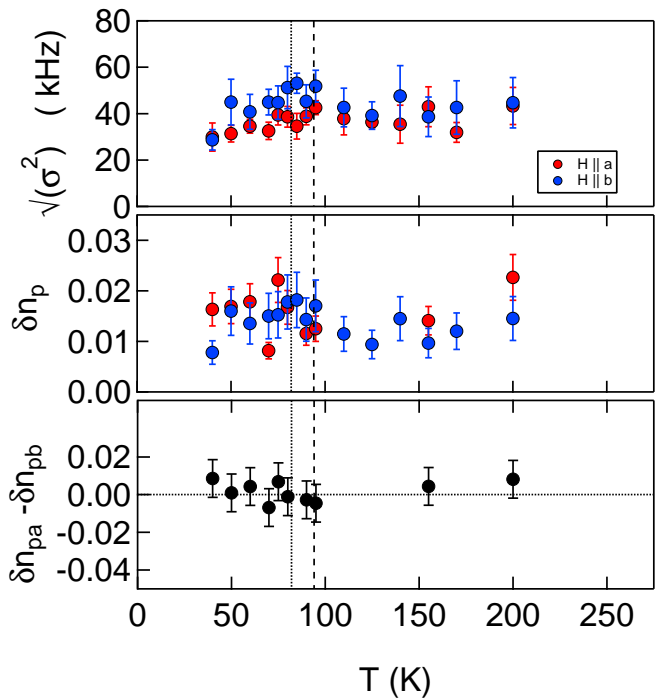}
 \caption{(Color online) (upper panel) The average of the rms second moments of the first satellite transitions ($\frac{3}{2}\leftrightarrow\frac{1}{2}$, $-\frac{3}{2}\leftrightarrow-\frac{1}{2}$) for the field parallel and perpendicular to the Cu-O bond axis. (middle panel) The rms second moment, $\delta n_p$ of the local hole concentration distribution of the two oxygen sites (see text for details), and (lower panel) the difference between $\delta n_{p,a}$ and $\delta n_{p,b}$ versus temperature. The dashed line indicates $T^* =94$ K and the dotted line indicates $T_c = 82$ K.}
\label{linewidths}
\end{center}
\end{figure}

The resonances in Fig. \ref{spectra} are broadened inhomogeneously. The fits to the spectra yield linewidths $\sigma_M = 36(8)$ kHz, $\sigma_Q = 98(6)$ kHz for the field in the plane, and $\sigma_M = 46(8)$ kHz, $\sigma_Q = 120(6)$ kHz for the field along $c$.  The dominant contribution to $\sigma_M$ arises from the inhomogeneity of the magnet (approximately $5\times 10^{-4}$ in the PPMS). Fig. \ref{linewidths} shows the temperature dependence of the linewidths of the first satellites ($\pm\frac{3}{2}\leftrightarrow\pm\frac{1}{2}$ transition) for the field along both $a$ and $b$.  The linewidth is slightly larger for the $b$ direction because  $K_b>K_a$  and therefore the inhomogeneity of the field translates into a greater linewidth.  At low temperatures the linewidths decrease slightly because $K$ is reduced. The quadrupolar contribution to the linewidth, $\sigma_Q$, reflects an inhomogeneous distribution of the local EFG at the O(1) sites. As discussed below, this distribution arises because of the intrinsic inhomogeneity in the hole occupations of the oxygen orbitals.

\section{Relaxation Measurements}

\begin{figure}
\begin{center}
 \includegraphics[width=\linewidth]{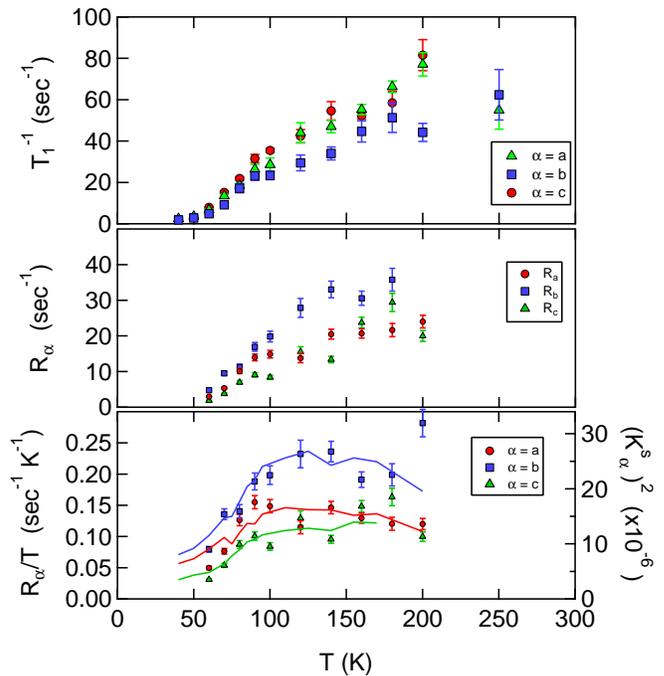}
 \caption{(Color online) (a) the \slrrtext, \slrr, for the field along each direction $a$ ($\bullet$, red), $b$ ($\blacksquare$, blue), and $c$ ($\blacktriangle$, green). (b) The relaxation rates $R_{\alpha}$ determined from the $T_{1\alpha}$ data as described in the text. (c) $R_{\alpha}/T$ and $(K^s)^2$ (solid lines) versus $T$.  Here $K^s = K -K_0$, where $K_0$ is determined in Fig. \ref{korringa}.}
\label{T1summary}
\end{center}
\end{figure}

In order to investigate the spin fluctuations present at the O(1) site, we have measured \slrr\ for all three orientations.  Since $\eta\neq 0$ for this site, the relationship between the spin fluctuation spectrum and the \slrrtext\ is complex.  The eigenstates, $|\phi_i\rangle$,  of the Hamiltonian (\ref{eqn:hamiltonian}) are superpositions of the $|m\rangle$ states, and transitions between these states are driven by fluctuations of the hyperfine field $\mathbf{h}(t)$.  This field arises through the interaction in Eq. \ref{eqn:milarice}, and gives rise to a fluctuating Hamiltonian:
\begin{equation}
\label{eqn:fluctuatinghyperfine}
\mathcal{H}_1(t) = \gamma\hbar \left[h_a(t)\hat{I}_a + h_b(t)\hat{I}_b + h_c(t)\hat{I}_c\right],
\end{equation}
where $\mathbf{h}(t) = g\mu_B\mathbb{C}\cdot\sum_i\mathbf{S}(\mathbf{r}+\mathbf{r}_i,t)$.
The spin fluctuations drive the nuclear spin lattice relaxation rates $R_{\alpha} = \gamma^2 \int_{-\infty}^{\infty}\overline{\langle h_{\alpha}(t)h_{\alpha}(t+\tau)\rangle} e^{-i\omega\tau}d\tau$ and we assume that  $\mathbf{h}(t)$ fluctuates independently in each direction. The detailed form of the magnetization relaxation and the dependence of \slrr\ on $R_{\alpha}$ are derived in the appendix.  The measured values are shown in Fig. \ref{T1summary} for all three directions.  The relaxation rates in each direction scale with one another and are reduced at low temperature due to the pseudogap. As seen in Fig. \ref{T1summary}(b), the fluctuations are largest along $b$, in agreement with the anisotropy of the hyperfine coupling.  Fig. \ref{T1summary}(c) displays $R_{\alpha}/T$ and $(K_{\alpha}^s)^2$ for each direction.  The scaling between these two quantities reflects Korringa relaxation ($T_1T \sim (K^s)^2$), in which the nuclei are relaxed by spin-flip scattering with quasiparticles, and agrees with previous studies. \cite{MMPT1inYBCO,shastrycuprates,MartindaleHammelYBCO,TakigawaBSCCOprl}  Fig. \ref{korringa} displays $(R_{\alpha}/T)^{1/2}$ versus $K$ with temperature as an implicit parameter.  The fact that the data are linear further supports this interpretation and implies that the temperature dependence of \slrr\ of the O(1) site is driven by a partial gapping of the density of states (pseudogap) rather than by spin fluctuations.

\section{Discussion}

\subsection{Scaling behavior}

\begin{figure}
\begin{center}
 \includegraphics[width=\linewidth]{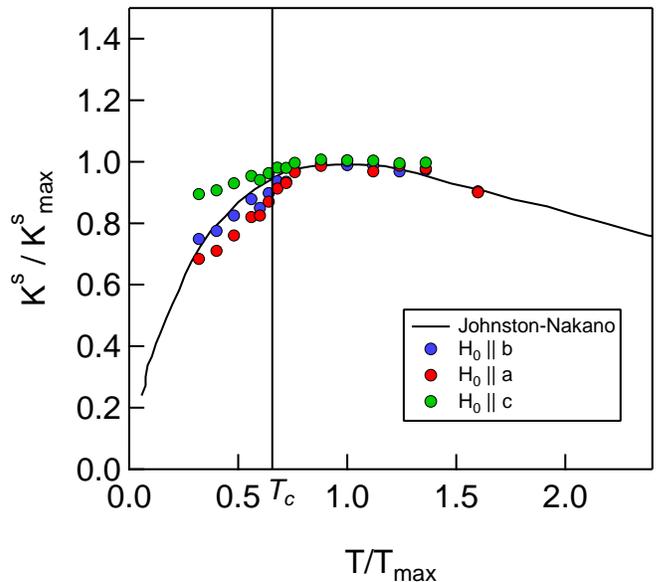}
 \caption{(Color online) The spin component of the Knight shift $K^s_{\alpha}$ versus $T/T^m$, where $T^m = 125$ K.  The solid line is the Johnston-Nakano scaling form \cite{BarzykinPinesTwoComponentReview}.}
\label{Kscaling}
\end{center}
\end{figure}

In recent years evidence has emerged that the the magnetic behavior of the \ybcox\ and \lscox\ cuprate families is best understood in terms of two electronic degrees of freedom. \cite{BarzykinPinesTwoComponentReview,HaaseSlichterTwoComponents2009}  Over a broad range of doping, the uniform susceptibility scales as $\chi(T,x) = \chi_0(x) + \chi^m\tilde{\chi}(T/T^m(x))$, where $\tilde{\chi}(T/T^m(x))$ is a universal function of $T/T_m$, and $T^m$ is the temperature where $\tilde{\chi}$ is maximum.  In Fig. \ref{Kscaling} we show the Knight shift scaled to fit this form for all three directions using $T^m=125$ K.  The data scale well for $T/T^m \gtrsim 0.75$, suggesting that $T^* \sim 94$ K.  Barzykin and Pines have found that $T^m = 1218 K (1-4.45p)$, where $p$ is the doping level in \lscox.  Using the measured value of $T^m$ we find $p = 0.201$, which is close to the value of $0.21$ based on the measured $T_c$.  These results confirm that the magnetic scaling in \bscco\ agrees quantitatively with that of \lscox, and implies that the magnetic scaling is uniform across all high $T_c$ families.

\subsection{Evidence for electronic inhomogeneity}

In recent years an increasing number of experiments on \bscco\ and other high temperature superconductors have uncovered evidence that the electronic degrees of freedom develop mesoscopic 1D stripe-like structures of inhomogeneous spin and charge densities below the pseudogap energy. \cite{Hinkov2007,Alldredge2008,mcElroySTMScience2005,Xu2009,Lawler2010,StanfordScienceBi2201PG2011}
STM measurements have revealed that the 90$^{\circ}$ $C_4$ rotation symmetry of the CuO$_2$ plane is broken by the electronic degrees of freedom in underdoped \bscco\ at the pseudogap energy, and that the two crystallographically equivalent planar O sites acquire different local hole concentrations.\cite{KohsakaScience2007}  In this case the local EFG and Knight shift should differ for the two sites, leading to a splitting or broadening of the NMR resonances.  These asymmetric patterns are short-ranged, may be dynamic, and may or may not couple to the external magnetic field used for the NMR experiments.  In order to investigate any asymmetry in the NMR response of the two planar O sites, we consider the linewidths of the parallel and perpendicular sites. There are three potential contributions to the linewidth: (\textit{i}) inhomogeneous fields from the magnet, ($ii$) a distribution of local Knight shifts, and ($iii$) a distribution of EFGs due to locally varying hole concentrations in the O 2p orbitals.  The field inhomogeneity $\beta_H = \delta H/H  \approx 5\times 10^{-4}$ is independent of direction and temperature, whereas the contribution from the Knight shift $\beta_K = \delta K/K$ depends on both quantities.  The distribution of local EFGs is directly related to the hole occupation in the O 2p orbitals via the relation $\nu_{\alpha\alpha} = \nu_{\alpha\alpha}^0 + q_{\alpha} n_p$, where $\nu_{\alpha\alpha}^0$ are material dependent constants,  $q_{\alpha} = 1.225$ MHz for $\alpha = (a,c)$, and $2.452$ MHz for $\alpha = b$, and  $n_p$ is the hole concentration in the O 2p orbital.\cite{HaaseHoleDensity}    Assuming the holes go exclusively into each of these two orbitals, $n_p = p/2  = 0.105$.  For a distribution $\delta n_p$ of local hole doping, the quadrupolar contribution to the linewidth would be $\sigma_Q = nq_{\alpha}\delta n_p$, where $n$ is the particular satellite transition.  If each of these contributions to the line broadening is independent of one another (which may not necessarily be the case (see Ref. \onlinecite{Haase2000})), then we can estimate the rms second moment of hole concentrations in both the parallel ($\alpha = b$) and perpendicular ($\alpha = a$) O 2p orbitals as:
\begin{equation}
\delta n_{p,\alpha}(T) = \frac{1}{nq_{\alpha}}\sqrt{\sigma^2_{\alpha}(T) - \nu_0^2(\beta_H^2 + \beta_K^2 K_{\alpha}^2(T))},
\end{equation}
where $\nu_0 = \gamma H_0$.  By fitting the central transition (for which $\sigma_Q=0$) at 110K we estimate $\beta_K = 0.0024$ for $\alpha = a$ and $0.0036$ for $\alpha=b$. We then calculate the temperature dependence of $\delta n_{p}(T)$ for both directions, as shown in the middle panel of Fig. \ref{linewidths}.  This result shows that the distribution of local hole dopings is roughly equal to 10-15\% of the average value, in agreement with previous STM estimates, and confirming that this inhomogeneity observed by STM is also present in the bulk.\cite{davisnatureinhomogeneity,mcElroySTMScience2005}

On the other hand, we find no evidence for any nematic asymmetry between the hole distribution for the parallel and perpendicular sites (lower panel in Fig. \ref{linewidths}).   Note, however, that the STM nematic asymmetry was observed in zero external field at finite bias in heavily underdoped samples. This NMR study was conducted in a field of 9 T in a slightly overdoped sample, and any dynamic nematic fluctuations would likely give rise to motional narrowing of the resonance.   Furthermore, it is unclear what role the magnetic field would have on the pattern of electronic excitations. Further studies in underdoped samples may help clarify these issues.

Surprisingly, there appears to be little or no contribution to the linewidth from the structural modulation.\cite{hirschfeldBSCCOmodulation} This modulation dramatically broadens the Cu and Bi resonances, as well as the oxygen site in the BiO layer.\cite{walstedtBSCCO,Bi209BSCCO,TakigawaBSCCOprl}  The CuO$_2$ planes are buffered from the disordered BiO layer by the SrO layers, and the relatively small quadrupolar moment of the oxygen render this site less sensitive to the structural modulation.

\subsection{Korringa relaxation}

\begin{figure}
\begin{center}
 \includegraphics[width=\linewidth]{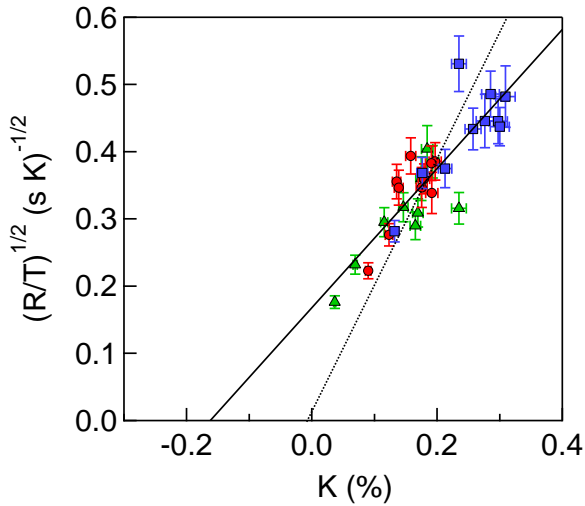}
 \caption{(Color online) $\sqrt{R_{\alpha}/T}$ versus $K_{\alpha}$ for $\alpha = a,b,c$.  The colors match those in Fig. \ref{spectra}.  The solid line is the best fit to the data as described in the text with slope 103(11) (s K)$^{1/2}$. The dotted line has a slope given by the Korringa value (see text).}
\label{korringa}
\end{center}
\end{figure}

The scaling between the Knight shift and the spin lattice relaxation rates $R_{\alpha}/T$  observed in Figs. \ref{T1summary} and \ref{korringa} is striking because it implies that the spin fluctuations responsible for the oxygen relaxation arise from quasiparticle scattering on the Fermi surface, rather than from  fluctuations of the Cu spins. In general, the nuclear \slrrtext\ is given by:
\begin{equation}
R_{\alpha} = \frac{k_BT}{2\mu_B^2\hbar^2}\lim_{\omega\rightarrow 0}\sum_{\mathbf q}F^2_{\alpha}(\mathbf{q})\frac{\chi"(\omega,\mathbf{q})}{\omega},
\end{equation}
where $\chi"(\omega, \mathbf{q})$ is the imaginary part of the electron spin susceptibility.\cite{MoriyaT1formula,CPSencyclopediaMagRes}
The hyperfine form factor $F_{\alpha}(\mathbf{q}) = 2C_{\alpha\alpha}\cos(q_{x,y}a/2)$ filters out Cu spin fluctuations at $\mathbf{Q}=(\pi/a,\pi/a)$ because the oxygen is located symmetrically between the two coppers. \cite{MilaRiceHamiltonian,shastrycuprates}   However recent neutron scattering studies on \bscco\ have confirmed the presence of incommensurate spin excitations at wavevectors $\mathbf{Q}_0\neq\mathbf{Q}$.\cite{Xu2009}  In this case the $F_{\alpha}(\mathbf{Q}_0)$ does not vanish, and spin fluctuations should contribute to the spin lattice relaxation rate. The temperature dependence of $R_{\alpha}$ would be driven by the correlation length, $\xi(T)$, and $R_{\alpha}/T$ would not scale with the shift $K_{\alpha}\sim\chi(\omega=0, \mathbf{q}=0)$.\cite{BarzykinCuprateScaling,ZhaBarzykinPines}  This scenario is inconsistent with the data.

On the other hand, for relaxation driven by quasiparticle scattering at the Fermi surface, $\sqrt{R_{\alpha}/T} = K_{\alpha}/\sqrt{2\kappa}$, where $\kappa=\mu_B^2/\pi k_B \hbar \gamma^2$ is the Korringa constant.\cite{CPSbook}    The best fit in Fig. \ref{korringa} yields a slope of 103(11) (s K)$^{-1/2}$, which is remarkably close to the theoretical value  187 (s K)$^{-1/2}$, suggesting that quasiparticle scattering rather than antiferromagnetic fluctuations is the dominant mechanism for the oxygen spin lattice relaxation. This result suggests that either (1) the oxygen is completely insensitive to the spin fluctuations, (2) there is a second electronic degree of freedom that couples to the oxygen, or (3) there or no antiferromagnetic spin fluctuations in the overdoped regime.  Since $\xi\sim a$ at $T=T_m$, it is likely that short range spin fluctuations continue to exist down to $T_c$.\cite{BarzykinCuprateScaling}

The discrepancy between the neutron scattering and the oxygen NMR data is well known, and others have proposed several possible explanations, including (1) the presence of another band or degree of freedom associated with the holes on the oxygen orbitals,\cite{WalstedtPRL1998,WalstedtCheongPRB2001,shastrycuprates} (2) a second transferred hyperfine coupling to the next nearest Cu sites,\cite{ZhaBarzykinPines} and (3) fluctuations of the EFG at the O site.\cite{Suter1248PRL2000} Another possibility is that the dynamical susceptibility contains two terms, one from the Fermi surface itself, and another from the incommensurate antiferromagnetic background.\cite{BarzykinCuprateScaling}  The incommensuration response implies a modulation in real space of the staggered magnetization with wavelength $\lambda \gg a$.\cite{Tranquada1995}  If the response were static, then there would be multiple oxygen sites with different hyperfine fields and relaxation rates.\cite{GrafeLESCO} The experiments clearly rule this static scenario out, thus we expect dynamic fluctuations of this modulated structure.  In this case the oxygen sites would be motionally narrowed, and all of the oxygen sites would experience antiferromagnetic fluctuations that would drive the spin lattice relaxation.

If this modulation contained higher order harmonics such that the real space spin density were a square wave rather than a sinusoid (discommensurations) the response would be commensurate with antiphase domain walls. \cite{ZhaBarzykinPines,BarzykinPinesTwoComponentReview} In this case the oxygen form factor would filter out the spin fluctuations, which resolves the apparent inconsistency between the INS and NMR results.\cite{ZhangRiceCuprates1988} In this case the relaxation of the oxygen may be driven by the dynamics of the domain walls. An alternative explanation is that the relaxation is driven by a second electronic degree of freedom. Recent NMR studies of the O and Cu shift in \lscox\ suggest the presence of a second spin component at the oxygen site.\cite{HaaseSlichterTwoComponents2009,BarzykinPinesTwoComponentReview} Which explanation hold for this case remains unclear at present.

\section{Summary}

We have measured the full anisotropy of the EFG, Knight shift, and spin lattice relaxation tensors at the planar oxygen site in an overdoped single crystal of \bscco. We present a complete analysis of the anisotropy of the spin fluctuations on the spin lattice relaxation rate for the oxygen. The Knight shift and \slrrtext\ reveal the presence of a pseudogap and the relaxation rate is well described by quasiparticle scattering at the Fermi surface. The Knight shift agrees with the scaling result implying that $T_m\sim 125$ K and $T^*\sim 94$ K.  Antiferromagnetic fluctuations of Cu spins do not appear to contribute to the oxygen relaxation rate, suggesting that the incommensurate order observed by neutron scattering must contain higher order harmonics.  The quadrupolar contribution to the resonance linewidth is consistent with a distribution of local hole concentrations in the oxygen 2p orbitals, in agreement with STM studies.  This result implies that the intrinsic electronic inhomogeneity is present in the bulk and is not purely a surface phenomenon.  On the other hand, the charge distribution in the O 2p orbitals appears to be identical for both the planar oxygen sites and remains temperature independent.  Further studies at lower dopings in the pseudogap phase may provide new information about the interplay of the local spin structure and the complex electronic patterns that emerge in the underdoped cuprates.

\appendix*
\section{}

Spin lattice relaxation is described by the master equation:
\begin{equation}
\label{eqn:master}
\frac{d\rho_{ii}(t)}{dt} = W_{ij}\left(\rho_{jj}(t)-\rho_{jj}0\right),
\end{equation}
where $\rho_{ii}$ are the diagonal components of the thermal averaged density matrix, $\hat{\rho}^0$ is the thermal equilibrium density matrix, and $W_{ij}$ is the spin lattice relaxation matrix.  The elements of $W_{ij}$ are given by:
\begin{equation}
W_{ij} = \frac{1}{\hbar^2}\left(J_{ijij} - \delta_{ij}\sum_k J_{kikj}\right),
\end{equation}
where
\begin{equation}
J_{ijkl} = \int_{-\infty}^{\infty}\overline{\langle \phi_i|\mathcal{H}_1(t)|\phi_j\rangle \langle \phi_l|\mathcal{H}_1(t+\tau)|\phi_k\rangle} e^{-i\omega\tau}d\tau
\end{equation}
and the overline signifies a thermal average.\cite{CPSbook}  If $[\hat{I}_z,\mathcal{H}]=0$ such that  $|\phi_i\rangle$ are eigenstates of $\hat{I}_z$, then the $J_{ijkl}$ vanish for any transition where $j\neq i\pm1$ and $k\neq l\pm1$ as per the usual selection rule.  For the O(1) site the $J_{ijkl}$ do not vanish because the eigenstates are superpositions of the $\hat{I}_z$ states.  Effectively this means that for the field along a direction $\alpha$, fluctuations of $\mathbf{h}(t)$ along \emph{all three} directions can give rise to transitions instead of just those two directions that are perpendicular to $\alpha$. The relaxation matrices, $\tilde{W}_{\alpha}$, depend simultaneously on $R_a$, $R_b$, and $R_c$. This simultaneous dependence means that the coupled differential equations in (\ref{eqn:master}) cannot be written in terms of a single effective time constant ($T_1$), and there is no closed form expression for the magnetization recovery function.

In order to surmount this difficulty, we expand the relaxation matrix as:
\begin{eqnarray}
\nonumber \tilde{W}_{\alpha}(R_a,R_b,R_c)&=&T_{1,\alpha}^{-1}(R_a,R_b,R_c)\tilde{w}_{\alpha}(R_a^0/R_b^0,R_a^0/R_c^0)+\\
&&\tilde{\delta W}_{\alpha}(R_a,R_b,R_c),
 \end{eqnarray}
where $\tilde{w}_{\alpha}(R_a^0/R_b^0,R_c^0/R_b^0)$ is a constant matrix, $T_{1,\alpha}^{-1}(R_a,R_b,R_c)$ is an overall time constant that depends on the relaxation components $R_{\alpha}$, and $\tilde{\delta W}_{\alpha}(R_a,R_b,R_c)$ is a small correction which we ignore. We estimate the ratios $R_a^0/R_b^0 = 0.329$ and $R_c^0/R_b^0=0.373$ as the squares of the hyperfine coupling ratios, since $R_\alpha\sim h_{\alpha}^2\sim C_{\alpha\alpha}^2$, where $C_{\alpha\alpha}$ is the hyperfine coupling. \cite{YoshinariYBCO,MilaRiceHamiltonian}  In this approximation, we assume a single component coupling model to isotropic Cu $S=1/2$ spins.  In the limit $\gamma H_0/\nu_{bb}\rightarrow\infty$ the eigenstates of (\ref{eqn:hamiltonian}) indeed commute with $\hat{I}_z$ and the relaxation matrices reduce to their conventional forms that depend only on a single time constant.  In our case $\gamma H_0/\nu_{bb} \approx 50$ so the correction term is small and the approximation is valid. In this case we find
\begin{widetext}
\begin{equation}
\tilde{w}_a\approx
\left(
\begin{array}{cccccc}
 -2.452 & 0 & 0 & 2.452 & 0 & 0.0001099 \\
 0 & -2.497 & 2.497 & 0 & 0.0001201 & 0 \\
 0 & 2.497 & -6.472 & 0 & 3.975 & 0.0002096 \\
 2.452 & 0 & 0 & -6.392 & 0.0002036 & 3.940 \\
 0 & 0.0001201 & 3.975 & 0.0002036 & -8.427 & 4.451 \\
 0.0001099 & 0 & 0.0002096 & 3.940 & 4.451 & -8.391
\end{array}
\right)
\end{equation}

\begin{equation}
\tilde{w}_b\approx
\left(
\begin{array}{cccccc}
 -2.488 & 0 & 0 & 2.488 & 0 & 0.00004005 \\
 0 & -2.487 & 2.487 & 0 & 0.00004574 & 0 \\
 0 & 2.487 & -6.466 & 0 & 3.979 & 0.00007869 \\
 2.488 & 0 & 0 & -6.469 & 0.00007528 & 3.980 \\
 0 & 0.00004574 & 3.979 & 0.00007528 & -8.456 & 4.477 \\
 0.00004005 & 0 & 0.00007869 & 3.980 & 4.477 & -8.458
\end{array}
\right)
\end{equation}

and
\begin{equation}
\tilde{w}_c\approx
\left(
\begin{array}{cccccc}
 -2.440 & 0 & 0 & 2.440 & 0 & 0.0002059 \\
 0 & -2.501 & 2.501 & 0 & 0.0002153 & 0 \\
 0 & 2.501 & -6.477 & 0 & 3.975 & 0.0003816 \\
 2.440 & 0 & 0 & -6.366 & 0.0003760 & 3.926 \\
 0 & 0.0002153 & 3.975 & 0.0003760 & -8.420 & 4.444 \\
 0.0002059 & 0 & 0.0003816 & 3.926 & 4.444 & -8.371
\end{array}
\right)\end{equation}

\end{widetext}
where
 \begin{eqnarray}
 T_{1a}^{-1} &=& R_b + 1.039 R_c\\
 T_{1b}^{-1} &=& R_a + 1.010 R_c\\
 T_{1c}^{-1} &=& R_b + 1.050 R_a.
\end{eqnarray}

The magnetization recovery measured along the $\alpha$ direction following an inversion pulse is then given by: $M_{\alpha}(t) = M_{0}(1 - 2f\phi_{\alpha}(t/T_{1\alpha}))$, where $\phi_{\alpha}(x) = \sum_{i=1}^5\lambda_i e^{-\beta_ix}$, and the constants $\lambda_i$ and $\beta_i$ are given   in Table \ref{tab:table1} for the first satellite transition of the O(1).

\begin{table}[b]
\caption{\label{tab:table1}%
Coefficients and exponents in the multiexponential spin lattice relaxation decay function for the O(1) nucleus.}
\begin{ruledtabular}
\begin{tabular}{rdddddd}
$i$ &
\multicolumn{1}{c}{$\lambda_a$}&
\multicolumn{1}{c}{$\beta_a$}&
\multicolumn{1}{c}{$\lambda_b$}&
\multicolumn{1}{c}{$\beta_b$}&
\multicolumn{1}{c}{$\lambda_c$}&
\multicolumn{1}{c}{$\beta_c$}\\
\colrule
1 & 0.454 & 14.84 & 0.446 & 14.92 & 0.457 & 14.82 \\
2 & 0.440 & 9.89 & 0.447 & 9.95 & 0.438 & 9.88 \\
3 & 0.0225 & 5.94 & 0.0251 & 5.97 & 0.0215 & 5.93 \\
4 & 0.0546 & 2.97 & 0.0535 & 2.98 & 0.0549 & 2.96 \\
5 & 0.0282 & 0.99 & 0.0286 & 0.99 & 0.0280 & 0.99 \\
\end{tabular}
\end{ruledtabular}
\end{table}

\begin{acknowledgements}
We thank P. Hirschfeld, S. Kivelson, D. Pines and C. P. Slichter for stimulating discussions. Work at UC Davis was supported by National Science Foundation under Grant No. DMR-1005393, and work at BNL is supported by the US Department of Energy (DOE) under Contract No. DE-AC02-98CH10886. N.C. thanks the Aspen Center for Physics for hospitality during which part of this manuscript was prepared.
\end{acknowledgements}

\bibliography{C:/Bibliography/CurroNMR}

\end{document}